\documentclass[a4paper]{article}

\usepackage{amsfonts}
\usepackage{amsmath}
\usepackage{graphicx}
\usepackage{latexsym}
\usepackage{lwa}
\usepackage{times}
\usepackage{url}
\usepackage{wrapfig}

\newcommand{\argmax}{\operatornamewithlimits{argmax}}

\title{Document Clustering Evaluation: Divergence from a Random Baseline}
\author{Christopher M. De Vries $^1$ \and Shlomo Geva $^1$
\and Andrew Trotman $^2$ \\
\vspace{-0.5em}\\
School of Electrical Engineering and Computer Science \\
Queensland University of Technology, Brisbane, Australia $^1$ \\
\vspace{-0.5em}\\
Department of Computer Science, University of Otago,
Dunedin, New Zealand $^2$ \\
\vspace{-0.5em}\\
{\em chris@de-vries.id.au \hspace{10px} s.geva@qut.edu.au \hspace{10px}
andrew@cs.otago.ac.nz}}

\begin{document}

\maketitle

\begin{abstract}
Divergence from a random baseline is a technique for the evaluation of document
clustering. It ensures cluster quality measures are performing work that
prevents ineffective clusterings from giving high scores to clusterings that
provide no useful result. These concepts are defined and analysed using
intrinsic and extrinsic approaches to the evaluation of document cluster
quality. This includes the classical clusters to categories approach and a novel
approach that uses ad hoc information retrieval. The divergence from a random
baseline approach is able to differentiate ineffective clusterings encountered
in the INEX XML Mining track. It also appears to perform a normalisation similar
to the Normalised Mutual Information (NMI) measure but it can be applied to any
measure of cluster quality. When it is applied to the intrinsic measure of
distortion as measured by RMSE, subtraction from a random baseline provides a
clear optimum that is not apparent otherwise. This approach can be applied to
any clustering evaluation. This paper describes its use in the context of
document clustering evaluation.
\end{abstract}

\section{Introduction}
This paper extends, motivates and analyses a document clustering evaluation
approach that compensates for ineffective document clusterings during
evaluation. An ineffective clustering is one that achieves a high score
according to a measure of document cluster quality but provides no value as a
clustering solution. Divergence from a random baseline is introduced and
formally defined to address ineffective clusterings in evaluation. A notion of
work performed by a clustering is introduced where ineffective cases appear to
perform no useful learning. The paper is concluded with a detailed analysis of
the results from the INEX 2010 XML Mining track. This paper clearly defines and
motivates this approach with theoretical and experimental analysis.

Ineffective document clusterings have been investigated using two extrinsic
evaluations. The first is the standard clusters to categories approach where
document clusters are compared to a ground truth set of category labels. The
second approach evaluates document clustering using ad hoc information retrieval
that has a use case for collection selection where a document collection is
distributed across many machines. A broker needs to direct a search query to
machines containing relevant documents. If the documents are allocated to
machines by document cluster, it is expected that only a few topical clusters
need to be searched. This is motivated by the cluster hypothesis
\cite{Jardine1971} that states relevant documents tend to be more similar to
each other than non-relevant documents. The Normalised Cumulative Cluster Gain
(NCCG) measure evaluates document clustering with respect to this use case for
ad hoc information retrieval.

The paper proceeds as follows. Section \ref{sec:inex} introduces the
collaborative XML document mining evaluation forum at INEX. Section
\ref{sec:document_clustering} introduces document clustering in an information
retrieval context and discusses different approaches. Evaluation of document
clustering using the clusters to categories approach and ad hoc relevance
judgements is discussed in Section \ref{sec:evaluation}. Sections
\ref{sec:ineffective}, \ref{sec:work} and \ref{sec:divergence} introduce and
define ineffective clusterings that perform no useful learning and can be
adjusted for by applying divergence from a random baseline. Section
\ref{sec:application} analyses the application of divergence from a random
baseline using the INEX 2010 XML mining track. The paper is concluded in Section
\ref{sec:conclusion}.

\section{INEX XML Mining Track}
\label{sec:inex}
The XML document mining track was run for six years at INEX, the Initiative for
the Evaluation of XML Information Retrieval
\footnote{\url{http://inex.otago.ac.nz/tracks/wiki-mine/wiki-mine.asp}}
\cite{Denoyer2007,Denoyer2008,Denoyer2009,Nayak2010,DeVries2011}. It explored
the emerging field of classification and clustering of semi-structured
documents.

Document clustering has been evaluated at INEX using the standard clusters to
categories approach, where categories extracted from the Wikipedia were used as
a ground truth. Clusterings produced by different systems were evaluated using
measures such as Purity, Entropy, F1 and NMI, indicating how well the clusters
match the categories.

A novel approach to document clustering evaluation was introduced at INEX in
2009 \cite{Nayak2010} and 2010 \cite{DeVries2011}. It used ad hoc information
retrieval to evaluate document clustering by using relevance judgments from
retrieval systems in the ad hoc track \cite{Tovar2010}. Ad hoc information
retrieval evaluation is a system based approach that evaluates how different
systems rank relevant documents. For systems to be compared, the same set of
information needs and documents have to be used. A test collection consists of
documents, statements of information need, and relevance judgments
\cite{Voorhees2002}. Relevance judgments are often binary and any document is
considered relevant if any of its contents can contribute to the satisfaction of
the specified information need. However, the ad hoc track at INEX provides
additional relevance information where assessors highlight the relevant text in
the documents. Information needs are also referred to as topics and contain a
textual description of the information need, including guidelines as to what may
or may not be considered relevant. Typically, only the keyword based query of a
topic is given to a retrieval system.

The ad hoc information retrieval based evaluation of document clustering is
motivated by the cluster hypothesis that suggests relevant documents are more
similar to each other than non-relevant documents; relevant documents tend to
cluster together. The spread of relevant documents over a clustering solution
was measured using the Normalised Cumulative Cluster Gain (NCCG) measure in the
INEX XML mining track in 2009 and 2010 \cite{Nayak2010,DeVries2011}. This
evaluation approach also has a specific use case in information retrieval. It
evaluates clustering of a document collection for collection selection.
Collection selection involves selecting a subset of a collection given a query.
Typically, these subsets are distributed on different machines. The goal is to
cluster documents such that only a small fraction of clusters, and therefore
machines, need to be searched to find most of the relevant documents for a given
query. This leads to improved run time performance as only a fraction of the
collection needs to be searched. The total load over a distributed system is
decreased as only a few machines need to be searched per query instead of every
machine. It also provides a clear use case for document clustering evaluation.
By contrast, comparing document clusters to predefined categories only
evaluates clustering as a match against a particular classification.

This paper uses the INEX 2010 XML Mining track dataset \cite{DeVries2011}. It is
a 146,225 document subset of the INEX XML Wikipedia collection determined by the
reference run used for the ad hoc track \cite{Arvola2011}. The reference run
contains the 1500 highest ranked documents for each of the queries in the ad hoc
track. The queries were searched using an implementation of Okapi BM25 in the
ATIRE \cite{Trotman2010} search engine.

Topical categories for documents are one of many views of extrinsic cluster
quality. They are derived from what humans perceive as topics in a document
collection. When categories are used for evaluation, a document clustering
system is given a score indicating how well the clusters match the predefined
categories. This is the most prevalent approach to evaluation of document
clustering in the research literature.

The categories for the INEX 2010 XML Mining collection were extracted from the
Wikipedia category graph which is noisy and nonsensical at times. Therefore, an
approach using shortest paths in the graph was used to extract 36 categories
\cite{DeVries2011}.

\section{Document Clustering}
\label{sec:document_clustering}
Document clustering is used in many different contexts, such as exploration of
structure in a document collection for knowledge discovery \cite{Tan1999},
dimensionality reduction for other tasks such as classification
\cite{Kyriakopoulou2007}, clustering of search results for an
alternative presentation to the ranked list \cite{Hearst1996} and
pseudo-relevance feedback in retrieval systems \cite{Lee2008}.

Recently there has been a trend towards exploiting semi-structured documents
\cite{Nayak2002,Denoyer2009}. This uses features such as the XML tree structure
and hyper-link graphs to derive data from documents to improve the quality of
clustering.

Document clustering groups documents into topics without any knowledge of the
category structure that exists in a document collection. All semantic
information is derived from the documents themselves. It is often referred to as
unsupervised clustering. In contrast, document classification is concerned with
the allocation of documents to predefined categories where there are labeled
examples to learn from. Clustering for classification is referred to as
supervised learning where a classifier is learned from labeled examples and
used to predict the classes of unseen documents.

The goal of clustering is to find structure in data to form groups. As a result,
there are many different models, learning algorithms, encoding of documents and
similarity measures. Many of these choices lead to different induction
principles \cite{Estivill-Castro2002} which result in discovery of different
clusters. An induction principle is an intuitive notion as to what constitutes
groups in data. For example, algorithms such as k-means \cite{Lloyd1982} and
Expectation Maximisation \cite{Dempster1977} use a representative based approach
to clustering where a prototype is found for each cluster. These prototypes are
referred to as means, centers, centroids, medians and medoids
\cite{Estivill-Castro2002}. A similarity measure is used to compare the
representatives to examples being clustered. These choices determine the
clusters discovered by a particular approach.

A popular model for learning with documents is the Vector Space Model (VSM)
\cite{Salton1975}. Each dimension in the vector space is associated with one
term in the collection. Term frequency statistics are collected by parsing the
document collection and counting how many times each term appears in each
document. This is supported by the distributional hypothesis \cite{Harris1954}
from linguistics that theorises that words that occur in the same context tend
to have similar meanings. If two documents use a similar vocabulary and have
similar  term frequency statistics then they are likely to be topically related.
The end result is a high dimensional, sparse document-by-term matrix who's
properties can be explained by Zipf distributions \cite{Zipf1949} in term
occurrence. The matrix represents a document collection where each row is a
document and each column is a term in the vocabulary. In the clustering process,
document vectors are often compared using the cosine similarity measure. The
cosine similarity measure has two properties that make it useful for comparing
documents. Document vectors are normalised to unit length when they are
compared. This normalisation is important since it accounts for the higher term
frequencies that are expected in longer documents. The inner product that is
used in computing the cosine similarity  has non-zero contributions only from
words that occur in both documents. Furthermore, sparse document representation
allows for efficient computation.

Different approaches exist to weight the term frequency statistics contained in
the document-by-term matrix. The goal of this weighting is to take into account
the relative importance of different terms, and thereby facilitate improved
performance in  common tasks such as classification, clustering and ad hoc
retrieval. Two popular approaches are TF-IDF \cite{Salton1988} and BM25
\cite{Robertson1995,Whissell2011}.

Clustering algorithms can be characterized by two properties. The first
determines if cluster membership is discrete. Hard clustering algorithms only
assign each document to one cluster. Soft clustering algorithms assign documents
to one or more clusters in varying degree of membership. The second determines
the structure of the clusters found as being either flat or hierarchical. Flat
clustering algorithms produce a fixed number of clusters with no relationships
between the clusters. Hierarchical approaches produce a tree of clusters,
starting with the broadest level clusters at the root and the narrowest at the
leaves.

K-means \cite{Lloyd1982} is one of the most popular learning algorithms for use
with document clustering and other clustering problems. It has been reported as
one of the top 10 algorithms in data mining \cite{Wu2008}. Despite research into
many other clustering algorithms it is often the primary choice for
practitioners due to its simplicity \cite{Guyon2009} and quick convergence
\cite{Arthur2009}. Other hierarchical clustering approaches such as repeated
bisecting k-means \cite{Steinbach2000}, K-tree \cite{DeVries2008} and
agglomerative hierarchical clustering \cite{Steinbach2000} have also been used.
Further methods such as graph partitioning algorithms \cite{Karypis1999}, matrix
factorisation \cite{Xu2003}, topic modeling \cite{Blei2003} and Gaussian
mixture models \cite{Dempster1977} have also been used.

The k-means algorithm \cite{Lloyd1982} uses the vector space model by
iteratively optimising $k$ centroid vectors which represent clusters. These
clusters are updated by taking the mean of the nearest neighbours of the
centroid. The algorithm proceeds to iteratively optimise the sum of squared
distances between the centroids and the set of vectors that they are nearest
neighbours to (clusters). This is achieved by iteratively updating the centroids
to the cluster means and reassigning nearest neighbours to form new clusters,
until convergence. The centroids are initialized by selecting $k$ vectors from
the document collection uniformly at random. It is well known that k-means is a
special case of Expectation Maximisation \cite{Dempster1977} with hard cluster
membership and isotropic Gaussian distributions.

The k-means algorithm has been shown to converge in a finite amount of time
\cite{Selim1984} as each iteration of the algorithm visits a possible
permutation without revisiting the same permutation twice, leading to a worst
case analysis of exponential time. Arthur et. al. \cite{Arthur2009} have
performed a smoothed analysis to explain the quick convergence of k-means
theoretically. This is the same analysis that has been applied to the simplex
algorithm, which has a $n^2$ worst case complexity but usually converges in
linear time on real data. While there are point sets that can force k-means to
visit every permutation, they rarely appear in practical data. Furthermore, most
practitioners limit the number of iterations k-means can run for, which results
in linear time complexity for the algorithm. While the original proof of
convergence applies to k-means using squared Euclidean distance
\cite{Selim1984}, newer results show that other similarity measures from the
Bregman divergence class of measures can be used with the same complexity
guarantees \cite{Banerjee2005}. This includes similarity measures such as
KL-divergence, logistic loss, Mahalanobis distance and Itakura-Saito distance.
Ding and He \cite{Ding2004} demonstrate the relationship between k-means and
Principle Component Analysis. PCA is usually thought of as a matrix
factorisation approach for dimensionality reduction where as k-means is
considered a clustering algorithm. It is shown that PCA provides a solution to
the relaxed k-means problem, thus formally creating a link between k-means and
matrix facortisation methods.

\section{Document Clustering Evaluation}
\label{sec:evaluation}
Evaluating document clustering is a difficult task. Intrinsic or internal
measures of quality such as distortion or log likelihood only indicate how well
an algorithm optimised a particular representation. Intrinsic comparisons are
inherently limited by the given representation and are not comparable between
different representations. Extrinsic or external measures of quality compare a
clustering to an external knowledge source such as a ground truth labeling of
the collection or ad hoc relevance judgments. This allows comparison between
different approaches. Extrinsic views of truth are created by humans and suffer
from the tendency for humans to interpret document topics differently. Whether a
document belongs to a particular topic or not can be subjective. To further
complicate the problem there are many valid ways to cluster a document
collection. It has been noted that clustering is ultimately in the eye of the
beholder \cite{Estivill-Castro2002}.

When comparing a cluster solution to a labeled ground truth, the standard
measures of Purity, Entropy, NMI and F1 are often used to determine the quality
of clusters with regard to the categories. Let $\omega=\{w_1,w_2,\ldots,w_{K}\}$
be the set of clusters for the document collection $D$ and
$\xi=\{c_1,c_2,\ldots,c_{J}\}$ be the set of categories. Each cluster and
category is a subset of the document collection, $\forall c \in \xi, w \in
\omega : c,w \subset D$. Purity assigns a score based on the fraction of a
cluster that is the majority category label, \begin{equation}\argmax_{c \in \xi}
\frac{|c \cap w_k|}{|w_k|},\end{equation} in the interval $[0,1]$ where 0 is
absence of purity and 1 is total purity. Entropy defines a probability for each
category and combines them to represent order within a cluster,
\begin{equation}-\frac{1}{\log J} \sum^{J}_{j=1} \frac{|c_j \cap w_k|}{|w_k|}
\log \frac{|c_j \cap w_k|}{|w_k|},\end{equation} which falls in the interval
$[0,1]$ where 0 is total order and 1 is complete disorder. F1 identifies a true
positive ($tp$) as two documents of the same category in the same cluster, a
true negative ($tn$) as two documents of different categories in different
clusters and a false negative ($fn$) as two documents of the same category in
different clusters where the score combines these classification judgements
using the harmonic mean, \begin{equation}\frac{2 \times tp}{2 \times tp + fn +
fp}.\end{equation} The Purity, Entropy and F1 scores assign a score to each
cluster which can be micro or macro averaged across all the clusters. The micro
average weights each cluster by its size, giving each document in the collection
equal importance in the final score. The macro average is simply the arithmetic
mean, ignoring the size of the clusters. NMI makes a trade-off between the
number of clusters and quality in an information theoretic sense. For a detailed
explanation of these measures please consult Manning et. al. \cite{Manning2008}.

\subsection{NCCG}
The NCCG evaluation measure has been used for the evaluation of document
clustering at INEX \cite{Nayak2010,DeVries2011}. It is motivated by van
Rijsbergen's cluster hypothesis \cite{Jardine1971}. If the hypothesis holds
true, then relevant documents will appear in a small number of clusters. A
document clustering solution can be evaluated by measuring the spread of
relevant documents for the given set of queries.

NCCG is calculated using manual result assessments from ad hoc retrieval
evaluation. Evaluations of ad hoc retrieval occur in forums such as INEX
\cite{Arvola2011}, CLEF \cite{Forner2011} and TREC \cite{Clarke2010}. The manual
query assessments are called the relevance judgments and have been used to
evaluate ad hoc retrieval of documents. The process involves defining a query
based on the information need, a retrieval system returning results for the
query and humans judging whether the results returned by a system are relevant
to the information need.

The NCCG measure tests a clustering solution to determine the quality of
clusters relative to the optimal collection selection. Collection selection
involves splitting a collection into subsets and recommending which subsets need
to be searched for a given query. This allows a retrieval system to search fewer
documents, resulting in improved runtime performance over searching the entire
collection. The NCCG measure has complete knowledge of which documents are
relevant to queries and orders clusters in descending order by the number of
relevant documents it contains. We call this measure an ``oracle'' because it
has complete knowledge of relevant documents. A working retrieval system does
not have this property, so this measure represents an upper bound on collection
selection performance.

Better clustering solutions in this context will tend to group together relevant
results for previously unseen ad hoc queries.  Real ad hoc retrieval queries and
their manual assessment results are utilised in this evaluation.  This approach
evaluates the clustering solutions relative to a very specific objective --
clustering a large document collection in an optimal manner in order to satisfy
queries while minimising the search space. The measure used for evaluating the
collection selection is called Normalised Cumulative Cluster Gain (NCCG)
\cite{Nayak2010}.

The Cumulative Gain of a Cluster (CCG) is defined by the number of relevant
documents in a cluster, $\textrm{CCG}(c,t)= \sum_{i=1}^{n}{Rel_i}$. A sorted
vector CG is created for a clustering solution, $c$, and a topic, $t$, where
each element represents the CCG of a cluster. It is normalised by the ideal gain
vector, \begin{equation}\textrm{SplitScore}(t,c)=\sum^{|\textrm{CG}|} \frac{
\textrm{cumsum} ( \textrm{CG} )}{n_{r}^2},\end{equation} where $n_{r}$ is total
number of relevant documents for the topic, $t$. The worst possible split places
one relevant document in each cluster represented by the vector CG1,
\begin{equation}\textrm{MinSplitScore}(t,c) = \sum^{|\textrm{CG1}|} \frac{
\textrm{cumsum}( \textrm{CG1} )}{n_{r}^2}.\end{equation} NCCG is calculated
using the previous functions, \begin{equation}\textrm{NCCG}(t,c)=
\frac{\textrm{SplitScore}(t,c) - \textrm{MinSplitScore}(t,c)} {1 -
\textrm{MinSplitScore}(t,c)}.\end{equation} It is then averaged across all
topics.

\subsection{Single and Multi Label Evaluation}
Both the clustering approaches and the ground truth can be single or multi
label. Examples of algorithms that produce multi label clusterings are soft or
fuzzy approaches such as fuzzy c-means \cite{Bezdek1984}, Latent Dirichlet
Allocation \cite{Blei2003} or Expectiation Maximisation \cite{Dempster1977}. A
ground truth is multi label if it allows more than one category label for each
document. Any combination of single or multi label clusterings or ground truths
are able to be used for evaluation. However, it is only reasonable to compare
approaches using the same combination of single or multi label clustering and
ground truths. Multi label approaches are less restrictive than single label
approaches as documents can exist in more than one category. There is redundancy
in the data whether it is clustering or a ground truth. This redundancy has a
real and physical costs when clustering is used for collection selection. More
storage and compute resources are required with a multi label clustering as one
document has to be stored and processed on more than one computer. A ground
truth can be considered a clustering and compared to another ground truth to
measure how well the ground truths fit each other. Furthermore, a ground truth
can be used as a clustering solution and used for collection selection.

The evaluation of document clustering using ad hoc information retrieval can be
viewed as being similar to an evaluation using a multi label category based
ground truth. A document can be relevant to more than one query. However, unlike
a category based approach, each query is evaluated separately and then averaged
across all queries. In contrast, all categories are evaluated at once and the
score is not averaged across categories.

\section{Ineffective Clustering}
\label{sec:ineffective}
In this paper we introduce the concept of an ineffective clustering. An
ineffective clustering produces a high score according to an evaluation measure
but does not represent any inherent value as a clustering solution.

The Purity evaluation measure has an obvious ineffective case. If each cluster
contains one document then it is 100\% pure with respect to the ground truth. A
single document is the majority of the cluster. As the goal of clustering is to
produce groups of documents or to summarise the collection, this is obviously
flawed as it does neither. The same applies to the Entropy measure as the
probability of a label for a cluster is 100\%, resulting in the highest possible
Entropy score.

The NCCG measure is ineffective when one cluster contains all the documents
except for every other cluster containing one document. The NCCG measure orders
clusters by the number of relevant documents they contain. A large cluster
containing most documents will almost always be ranked first. Therefore, almost
all relevant documents will exist in one cluster, achieving almost the highest
score possible.

\section{Work Performed by a Clustering}
\label{sec:work}
To overcome ineffective clusterings in the previous section, we introduce the
concept of work performed by a clustering approach. Work is defined as an
increase in quality of a clustering over a simple approach that ignores the
documents being clustered. A useful clustering performs work beyond an approach
that is purely random and ignores document content. If a random approach that
performs no useful learning performs equally to an approach that attempts to
learn from that data, it would appear that nothing has been achieved by
analysing the data. We suggest that an ineffective clustering performs no useful
learning. This is supported by a theoretical and experimental analysis in the
following sections.

\begin{figure}[!ht]
\centering
\includegraphics[width=0.3\textwidth]{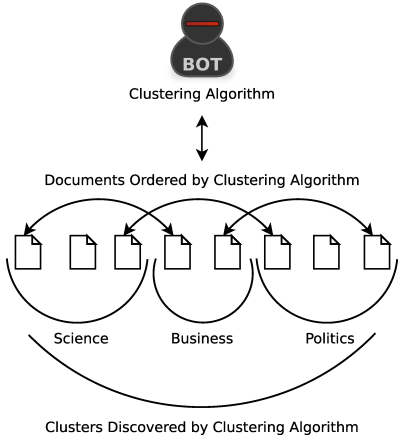}
\caption{A Clustering Produced by a Clustering Algorithm}
\label{fig:clustering_algorithm}
\end{figure}

\begin{figure}[!ht]
\centering
\includegraphics[width=0.3\textwidth]{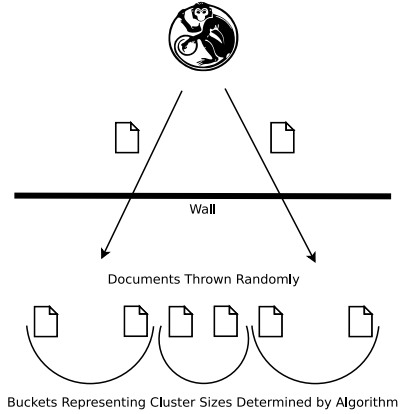}
\caption{A Random Baseline Distributing Documents into Buckets the Same Size as
a Clustering}
\label{fig:monkey_baseline}
\end{figure}

Figures \ref{fig:clustering_algorithm} and \ref{fig:monkey_baseline} illustrate
an approach using a clustering algorithm and a random approach that ignores
document content. The difference in cluster quality between these two approaches
represents work completed by a clustering algorithm.

\section{Divergence from a Random Baseline}
\label{sec:divergence}
Many measures of cluster quality can give high quality scores for particular
clustering solutions that are not of high quality by changing the number of
clusters or number of documents in each cluster.

Measures that can be misled by creating an ineffective clustering can be
adjusted by subtraction from a randomly generated clustering with the same
number of clusters with the same number of documents in each cluster. Figures
\ref{fig:clustering_algorithm} and \ref{fig:monkey_baseline} highlight this
example where the random baseline distributes documents into buckets the same
size as the clusters found by the clustering algorithm. Apart from the random
assignment of documents to clusters, the random baseline appears the same as the
real solution. Therefore, each clustering evaluated requires a random baseline
that is specific to that clustering. The baseline is created by shuffling the
documents uniformly randomly and splitting them into clusters the same size as
the clustering being measured. The score for the random baseline clustering is
subtracted from the matching clustering being measured.

The divergence from a random baseline approach can be applied to any measure
of cluster quality whether it is intrinsic of extrinsic. However, it does
require an existing measure of cluster quality. It is not a measure by itself
but an approach to ensure a clustering is doing something sensible. Although we
have highlighted its use for document clustering evaluation, it can be used for
any clustering evaluation.

There are two issues at play here. Firstly, different distributions of cluster
sizes can lead to arbitrarily high scores. The second issue is determining if
the clustering algorithm is effectively learning with respect to a measure of
quality. The divergence from a random baseline takes care of ineffective
solutions in either case. If the internal ordering of clusters is no better than
random noise then it achieves a score of zero. A negative score could be
achieved as the random baseline scores a positive value using most measures on
most data sets. It is possible for a clustering to have a worse score than the
baseline. For example, a clustering approach could maximise dissimilarity of
documents in clusters. This will create a solution where the most dissimilar
documents are placed together, resulting in a worse score than random
assignment. The random assignment does not bias the clustering towards or away
from the measure of quality. If a clustering approach is in fact learning
something with respect to the measure of quality, then it is expected that is
will be biased towards it. Alternatively, if we reverse the optimisation
process, it should be biased away from it.

Let $\omega=\{w_1,w_2,\ldots,w_{K}\}$ be the set of clusters for the document
collection $D$ and $\xi=\{c_1,c_2,\ldots,c_{J}\}$ be the set of categories. Each
cluster and category is a subset of the document collection, $\forall c \in \xi,
w \in \omega : c,w \subset D$. We define the probability of a category in the
baseline given a cluster as, $P_b(c_j | w_k) = \frac{|c_j|}{\sum_i |c_i|}$. The
probability of a category given a cluster in the baseline only depends on the
size of the categories. The baseline is a uniformly randomly shuffled list of
documents that has been split into clusters that match the cluster size
distribution in the solution being evaluated. Thus, within each cluster in the
baseline is random uniform noise. It is not biased by the document
representation. So, it is expected categories will occur at a rate proportional
to the category's size. For example, if there are three categories $A, B, C$
containing $10, 20, 30$ documents, each cluster in the baseline is expected to
contain approximately $\frac{10}{60} A, \frac{20}{60} B, \textrm{and}
\frac{30}{60} C$. This only reflects the size distribution of the categories.

We let any measure of a cluster quality be interpreted as a probability.
Although this is not formally the case for all measures, it serves as a
reasonable explanation. We define the probability of a category in a cluster
given the ground truth as, $P_s(c_j | w_k) = \textrm{any measure of cluster
quality}$.

The Purity measure assigns an actual probability to each cluster when there is a
single label ground truth. All the probabilities combined accumulate to one,
$\sum_j \frac{|c_j \cap w_k|}{|w_k|} = 1$, and the category with the largest
maximum likelihood estimate is assigned to each cluster, $P_\textrm{Purity}(c_j
| w_k) = \argmax_{c_j} \frac{|c_j \cap w_k|}{|w_k|}$. This is the proportion of
the cluster that has the majority category label. It also represents the same
process of using clustering for classification with labeled data where an
unseen sample is labeled based on the majority category label of the cluster it
is nearest neighbour to. We define $d$ as a document in $D$. The ground truth is
restricted to being single label where a document, $d$, only has only one label
in one category in the ground truth, $\forall d \in D, c_i \in \xi, c_j \in \xi
: d \in c_i \land d \notin c_j \land c_i \neq c_j$.

The adjusted measure is the difference between the submission and the baseline.
We define the adjusted probability of a category given a cluster as,
$P_a(c_j |w_k) = P_s(c_j | w_k) - P_b(c_j | w_k)$.

An alternative formal view of divergence from a random baseline can be defined
by a quality function, $m : \mathbb{PP(Z \times Z)} \rightarrow \mathbb{R}$,
that takes a set of clusters as a set of set of (document, category label)
pairs, $s$, and returns a real number indicating the quality of the clustering.
Examples of these cluster quality functions are Entropy, F1, NCCG, Negentropy,
NMI and Purity. There exists a function, $r : \mathbb{PP(Z \times Z)}
\rightarrow \mathbb{PP(Z\times Z)}$, that generates a random baseline, $b$,
given a clustering solution, $s$. The baseline has the same number of clusters
as the clustering solution, $|b| = |s|$. For every cluster in each of the
original clustering, $s$, and the baseline, $b$, the corresponding clusters
contain the same number of documents, $\forall k : |s_k| = |b_k|$. The adjusted
measure, $m_a : \mathbb{PP(Z \times Z)} \rightarrow \mathbb{R}$, becomes,
$m_a(s) = m(s) - m(r(s))$.

\section{Application at the INEX 2010 XML Mining Track}
\label{sec:application}
Participants were asked to submit multiple clustering solutions containing
approximately 50, 100, 200, 500 and 1000 clusters. The categories extracted
contained 36 categories due to only using categories with greater than 3000
documents. This choice was arbitrary and the decision for cluster sizes was made
based on the number of documents in the collection before the categories were
extracted. The number of categories in a document collection is subjective.
Therefore, a direct comparison of 36 clusters with 36 categories is not
necessary. Measuring how the categories behave over multiple cluster sizes
indicates the quality of clusters and the trend can be visualised.

\begin{wrapfigure}{R}{0.5\columnwidth}
\centering
\includegraphics[width=0.5\columnwidth]{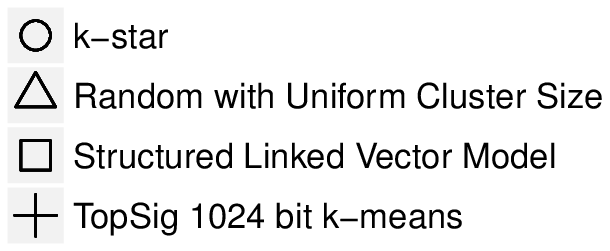}
\caption{Legend}
\label{fig:legend}
\end{wrapfigure}

\begin{figure}[!ht]
\centering
\includegraphics[width=0.45\textwidth]{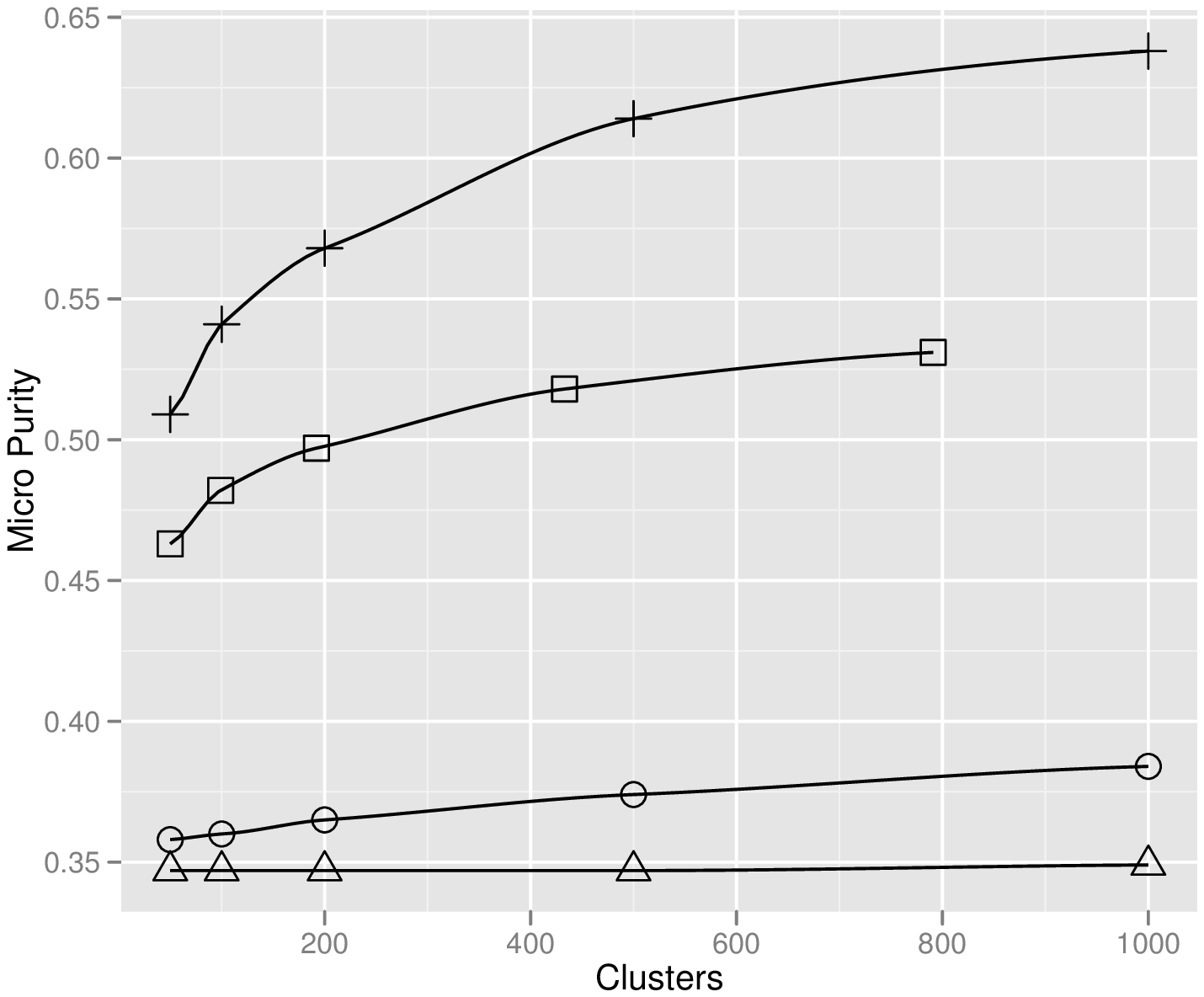}
\caption{Purity}
\label{fig:Purity}
\end{figure}

\begin{figure}[!ht]
\centering
\includegraphics[width=0.45\textwidth]{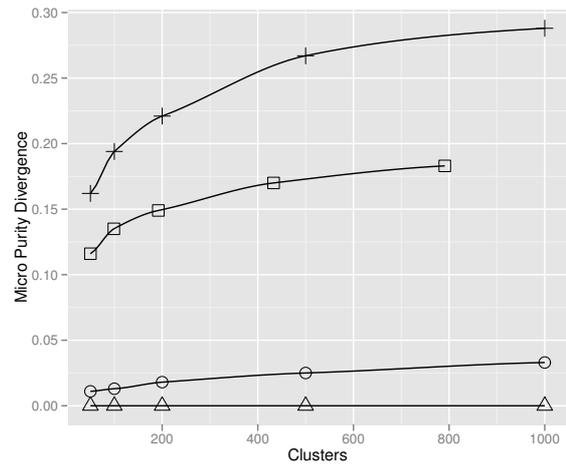}
\caption{Purity Subtracted from a Random Baseline}
\label{fig:adjusted_Purity}
\end{figure}

\begin{figure}[!ht]
\centering
\includegraphics[width=0.45\textwidth]{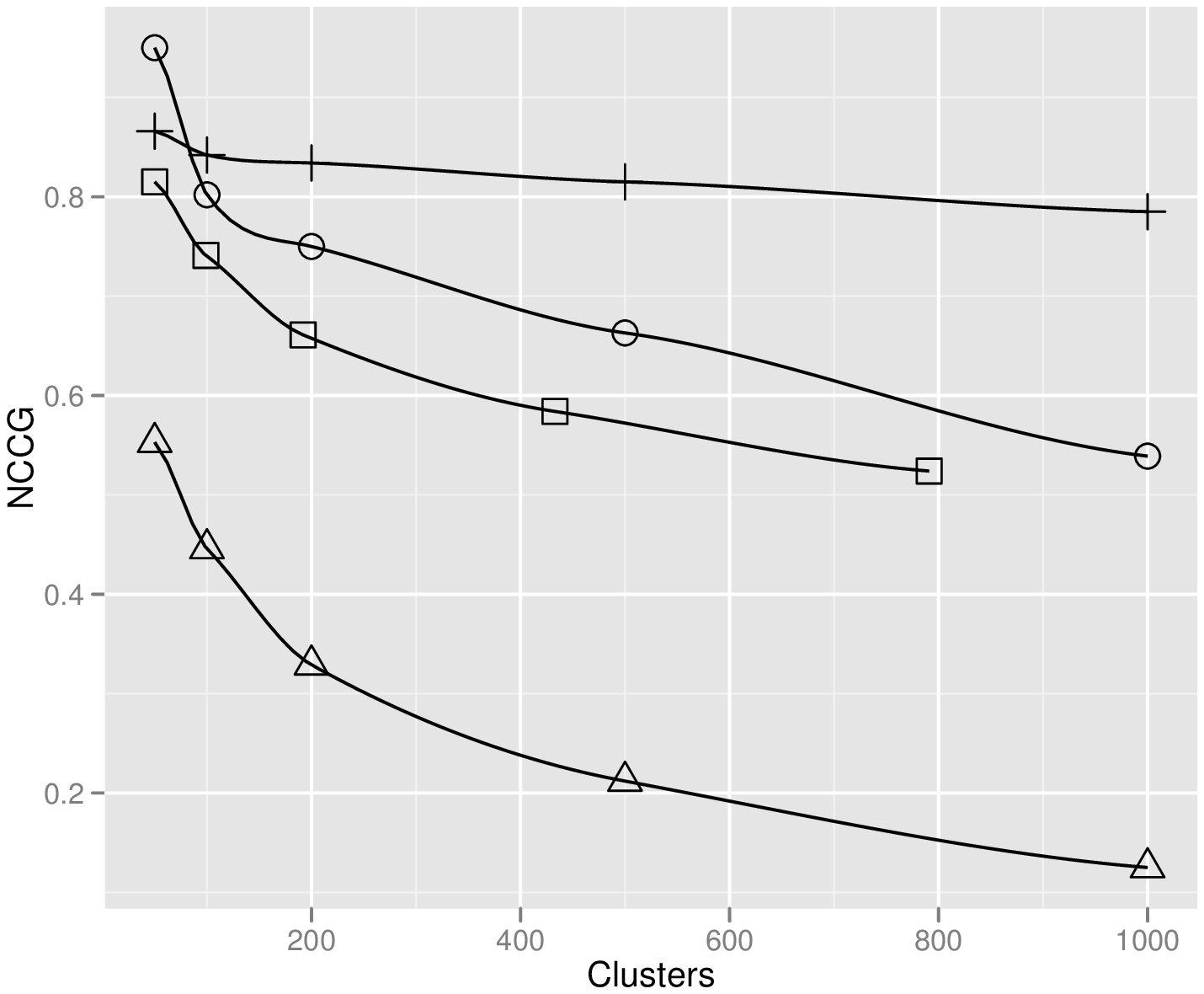}
\caption{NCCG}
\label{fig:NCCG}
\end{figure}

\begin{figure}[!ht]
\centering
\includegraphics[width=0.45\textwidth]{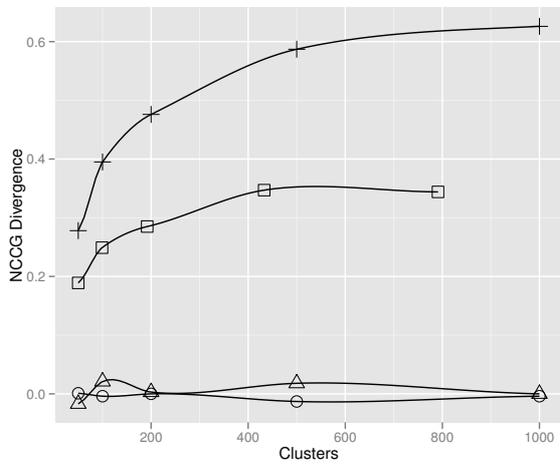}
\caption{NCCG Subtracted from a Random Baseline}
\label{fig:adjusted_NCCG}
\end{figure}

\begin{figure}[!ht]
\centering
\includegraphics[width=0.45\textwidth]{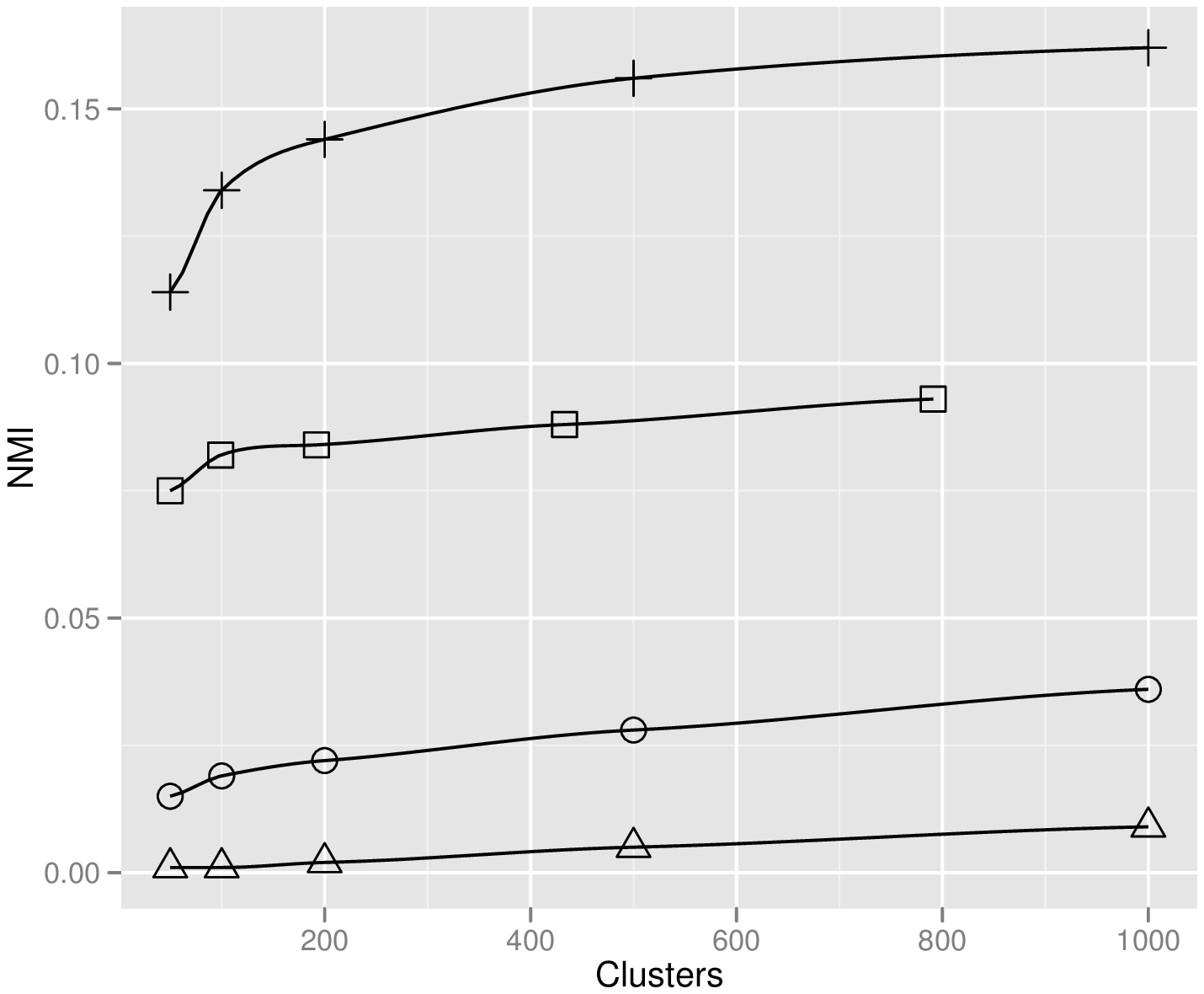}
\caption{NMI}
\label{fig:NMI}
\end{figure}

\begin{figure}[!ht]
\centering
\includegraphics[width=0.45\textwidth]{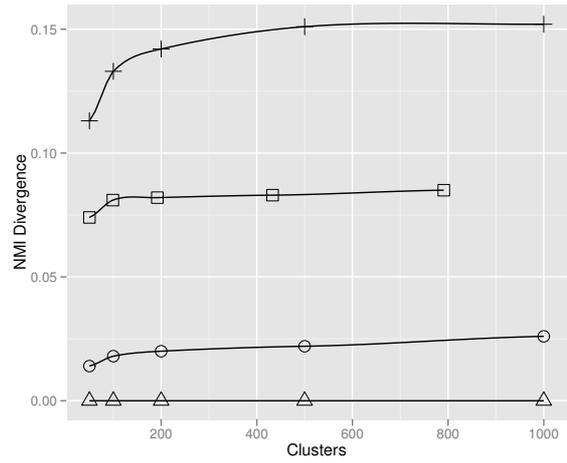}
\caption{NMI Subtracted from a Random Baseline}
\label{fig:adjusted_NMI}
\end{figure}

A legend for Figures \ref{fig:Purity} to \ref{fig:adjusted_NMI} can be found in
Figure \ref{fig:legend}. The Structured Linked Vector Model (SLVM)
\cite{Wang2010} incorporates document structure, links and content. The k-star
\cite{Tovar2010} is an iterative clustering method for grouping documents. The
TopSig approach \cite{Geva2011} produces binary strings that represent documents
and a modified k-means algorithm that works directly with this representation.

Submissions using the k-star method at INEX 2010 \cite{Tovar2010} contained
several large clusters and many other small clusters. This exposed weakness in
the NCCG measure, which resulted in inappropriately high scores. When the scores
are subtracted from a random baseline with the same properties they performed no
better than a randomly generated solution. This can be clearly seen in Figures
\ref{fig:NCCG} and \ref{fig:adjusted_NCCG} where the k-star method changes
drastically between the original score and the score when subtracted from a
random baseline.

The NMI measure is almost unaffected by subtraction from a random baseline where
as other measures have a larger difference. Figures \ref{fig:NMI} and
\ref{fig:adjusted_NMI} highlight this property on submissions from INEX 2010.
This suggests that the normalisation we have proposed is similar to that of NMI
but is applicable to any measure of cluster quality whether it is intrinsic or
extrinsic. Figures \ref{fig:Purity} and \ref{fig:adjusted_Purity} demonstrate
how the difference between the adjusted and unadjusted measures is larger for
measures that are not normalised. Each line represents a different document
clustering system. The bottom most line in each graph is a randomly generated
clustering submission where a category for a document is selected uniformly at
random from the set of categories. Note that this random clustering in the
figures differs from the random baseline. The cluster size distribution is also
uniform. A random baseline has a cluster size distribution that is specific to
the clustering being evaluated. When compared to the random baseline the
expected results are achieved, with a score of zero for all cluster sizes. Note
that without adjusting the cluster size distribution, it is not able to
differentiate ineffective clusterings as per the NCCG metric in Figure
\ref{fig:adjusted_NCCG}. Subtracting the random submission with uniform cluster
sizes from the NCCG submission does not reduce its score to zero as can be seen
in Figure \ref{fig:NCCG}.

\begin{figure}[!ht]
\centering
\includegraphics[width=0.45\textwidth]{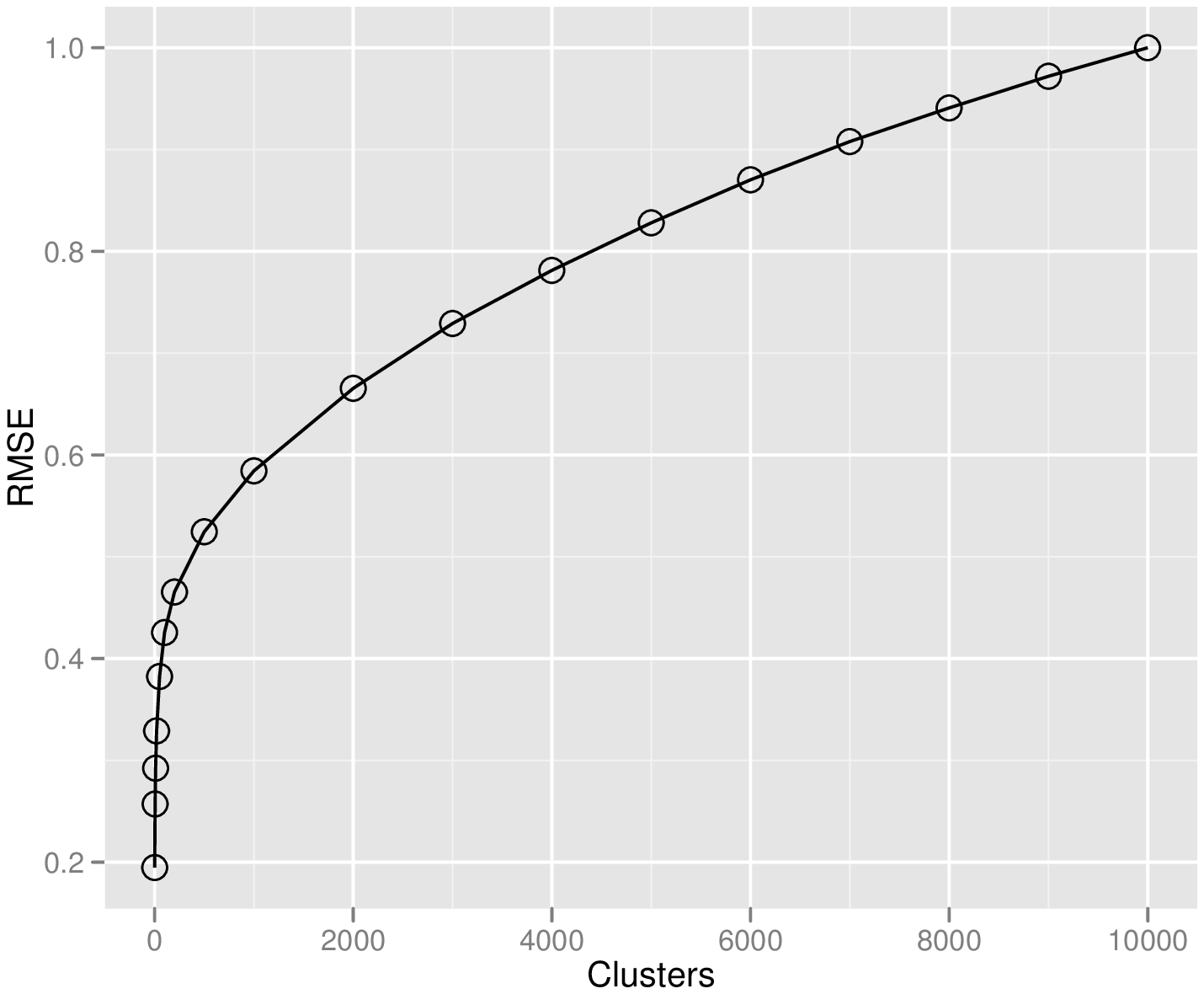}
\caption{RMSE}
\label{fig:RMSE}
\end{figure}

\begin{figure}[!ht]
\centering
\includegraphics[width=0.45\textwidth]{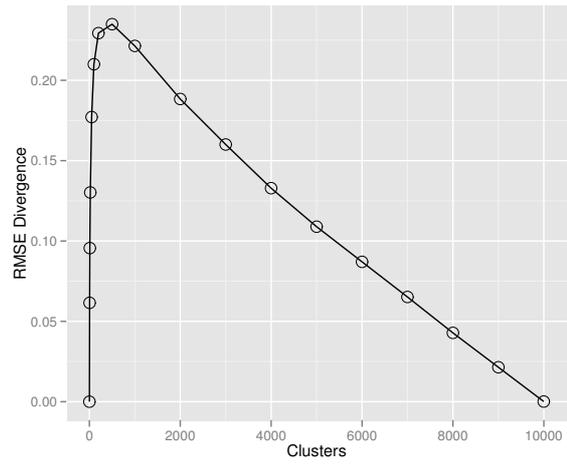}
\caption{RMSE Subtracted from a Random Baseline}
\label{fig:adjusted_RMSE}
\end{figure}

Figures \ref{fig:RMSE} and \ref{fig:adjusted_RMSE} demonstrate the application
of the divergence from random baseline approach on an intrinsic measure. RMSE is
the Root Mean Squared Error of the clustering using the cosine similarity
measure. The higher value the better the clustering. A cosine similarity of 1
indicates the document and the cluster centre are identical. A score of 0
indicates they are orthogonal and therefore have no overlap in vocabulary. This
experiment was run on a 10,000 document randomly selected sample. The k-means
algorithm was used to produce $k$ clusters between 1 and 10,000. Subtraction
from a random baseline assigns a score of zero to these ineffective cases.
Furthermore, it provides a clear maximum for RMSE.

\section{Conclusion}
\label{sec:conclusion}
In this paper we introduced problems encountered in evaluation of document
clustering. This is the concept of ineffective clustering and a notion of
work. The divergence from random baseline approach deals with these corner
cases and increases the confidence that a clustering approach is achieving
meaningful learning with respect to any view of cluster quality. It is also
applicable to any clustering evaluation but was only discussed in the context
of document clustering in this paper.

Divergence from a random baseline was formally defined and analysed
experimentally with both intrinsic and extrinsic measures of cluster quality.
Furthermore, this approach appears to be performing a normalisation similar to
that performed by NMI. It also provides a clear optimum for distortion as
measured by RMSE.

\bibliographystyle{plain}
\small{\bibliography{divergence_baseline}}

\end{document}